\documentclass{Interspeech2024}
\usepackage{authblk}
\usepackage{multirow}
\usepackage{graphicx}
\usepackage{subcaption}
\interspeechcameraready

\title{Serialized Output Training by Learned Dominance}

\name[affiliation={1}]{Ying}{Shi}
\name[affiliation={3}]{Lantian}{Li}
\name[affiliation={4}]{Shi}{Yin}
\name[affiliation={2}]{Dong}{Wang}
\name[affiliation={1}]{Jiqing}{Han}

\address{
  $^1$School of Computer Science and Technology, Harbin Institute of Technology, China \\
  $^2$Center for Speech and Language Technologies, BNRist, Tsinghua University, China \\
  $^3$School of Artificial Intelligence, Beijing University of Posts and Telecommunications, China \\
  $^4$Huawei Technologies Co., Ltd. China \\
 \thanks{
 D.W. (wangdong99@mails.tsinghua.edu.cn) and J.H. (jqhan@hit.edu.cn) are the corresponding authors.
 }  
}
\email{$^{2}$wangdong99@mails.tsinghua.edu.cn, $^{1}$jqhan@hit.edu.cn}


\begin{document}

\maketitle

\begin{abstract}
    

Serialized Output Training (SOT) has showcased state-of-the-art performance in multi-talker speech recognition by sequentially decoding the speech of individual speakers. 
To address the challenging label-permutation issue, prior methods have relied on either the Permutation Invariant Training (PIT) or the time-based First-In-First-Out (FIFO) rule. 
This study presents a model-based serialization strategy that incorporates an auxiliary module into the Attention Encoder-Decoder architecture, 
autonomously identifying the crucial factors to order the output sequence of the speech components in multi-talker speech. 
Experiments conducted on the LibriSpeech and LibriMix databases reveal that our approach significantly outperforms the PIT and FIFO baselines in both 2-mix and 3-mix scenarios. 
Further analysis shows that the serialization module identifies dominant speech components in a mixture by factors including loudness and gender, and orders speech components based on the dominance score. 

\end{abstract}

\section{Introduction}
\label{sec:intro}

Automatic Speech Recognition (ASR) technology has experienced rapid advancements in recent years, attributed in part to the introduction of 
end-to-end architectures~\cite{chorowski2015attention,chan2016listen,zeyer2019comparison} based on state-of-the-art network structures like 
Transformers~\cite{karita2019comparative,wang2020transformer} and Conformers~\cite{gulati2020conformer,zeineldeen2022conformer,zhang2020pushing}. 
For example, the Conformer model~\cite{zhang2020pushing} has achieved a Word Error Rate (WER) below 2\% on the Librispeech benchmark
\footnote[1]{https://paperswithcode.com/sota/speech-recognition-on-librispeech-test-clean}. 
However, it is important to note that such low WERs are attained in the context of single-talker speech. 
In scenarios where multiple speakers talk simultaneously, single-talker systems encounter confusion in determining which speaker to focus on. 
Multi-talker speech recognition sets a more ambitious objective: transcribing the speech of \textit{\textbf{all}} speakers, even when their 
signals overlap~\cite{wu2021investigation,settle2018end,masumura2023end,kang2024cross}.

Early research adopted a multi-head design, where each head produces transcriptions for one speaker within 
the overlapped speech~\cite{weng2015deep,lu2021streaming,yu2017permutation}. A key challenge of this architecture is the uncertainty in 
label assignment, where it remains unclear which speaker's text labels should be assigned to each head during loss computation. 
While some studies address this uncertainty by leveraging biased information such as energy~\cite{weng2015deep} or starting time~\cite{lu2021streaming}, 
Permutation Invariant Training (PIT) has gained widespread popularity~\cite{yu2017permutation,qian2018single,meng2023sidecar,zhang2020improving}.

One significant drawback of the aforementioned multi-head architecture is the necessity to pre-define the number of speakers, with larger speaker counts leading to 
increased computational demands. Serialized Output Training (SOT)~\cite{kanda2020serialized,kanda2021large}, built on the Attention Encoder-Decoder (AED) 
architecture~\cite{chorowski2014end}, addresses these challenges by sequentially outputting the transcriptions of each speaker. 
SOT utilizes a unified decoder for transcribing all speakers' speech, eliminating the need for multiple heads. 
However, determining the sequence of target text labels, known as \emph{label serialization}, remains crucial, akin to label assignment in the multi-head architecture. 
Two prevalent serialization strategies are First-In-First-Out (FIFO)~\cite{tripathi2020end} and sequential PIT~\cite{kanda2020serialized,kanda2020joint}. FIFO concatenates 
labels of single-talker speech speaker by speaker, ordered according to starting time, while sequential PIT orders the labels of different speakers 
in a manner that results in minimal loss based on the output of the current model. Neither FIFO nor PIT is flawless. 
Sequential PIT employs an `indeterministic' strategy, lacking a guarantee for consistent logic in serialization — a prerequisite for 
effective model learning from serialized labels. In contrast, FIFO is inherently deterministic and theoretically superior to PIT, which has 
been supported by the findings from Kanda et al.~\cite{kanda2020serialized}. 
However, when the starting time bias is ambiguous, uncertainty affects both training and inference, resulting in a significant performance drop, as demonstrated in the experimental section.

\begin{figure*}[ht!]
     \centering
     \vspace{-2mm}
     \includegraphics[width=0.88\linewidth, height=0.25\linewidth]{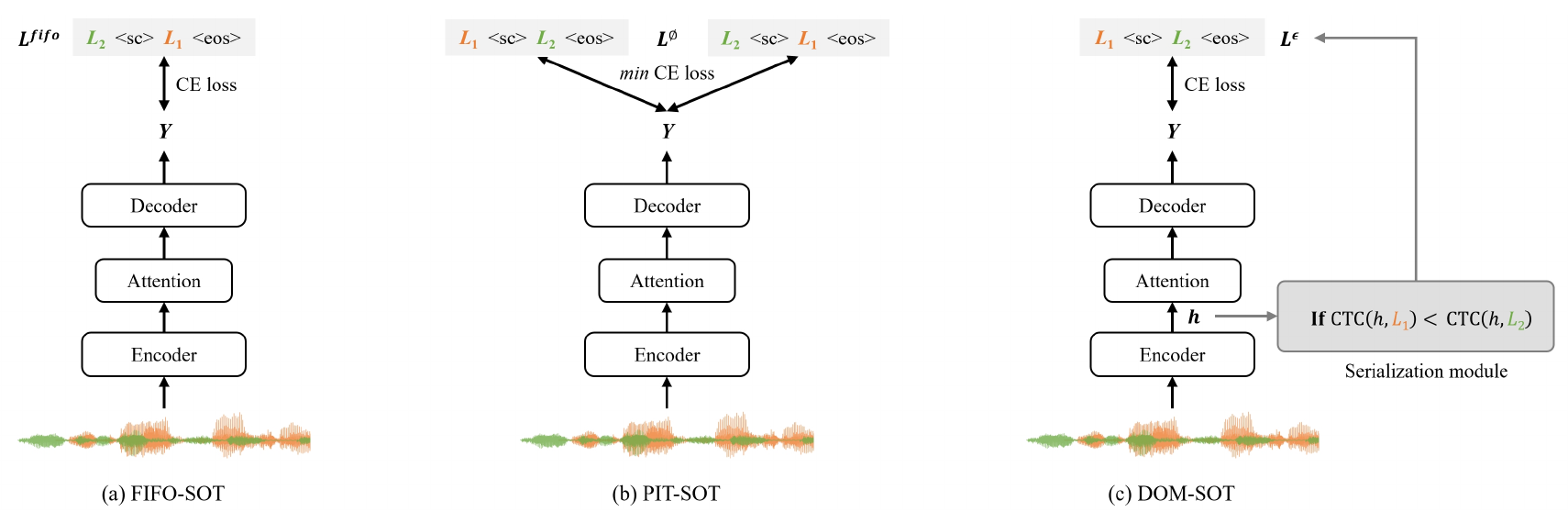}
     \caption{Diagram of FIFO-SOT, PIT-SOT, and our DOM-SOT}
     \label{fig:diag}
     \vspace{-2mm}
\end{figure*}

In this paper, we present a novel serialization strategy for multi-talker ASR with SOT. 
The fundamental concept involves training an explicit serialization module to dynamically determine the label order, departing from predefined biases or 
the minimum-loss order associated with PIT. We hope that the model can learn to discover factors by itself and use these factors to order the speech 
components, akin to human decision-making on which speech component to focus first. Specifically, 
the serialization module comprises a connectionist temporal classification (CTC) decoder~\cite{graves2012connectionist} integrated with 
the encoder of the AED architecture, trained by minimizing the \emph{minimum} CTC loss between 
the encoder output and labels of each speech component. This minimum CTC loss enforces the serialization module to identify a `dominant component' 
from the mixed speech to recognize, akin to humans unconsciously paying attention to the dominant speech in 
a cocktail party environment~\cite{kawata2020neural,mast2002dominance,jie2010recognize}. Once the serialization module is fully trained, the CTC loss 
assigned to each speech component, which can be regarded as a \emph{dominance score}, is used to order the speech components in mixed signals. 
We, therefore, name the new approach \textbf{dominance-based serialization}.

Our contribution is twofold: (1) We present a novel dominance-based serialization strategy for multi-talker ASR with SOT. 
Experimental results demonstrated that this strategy outperforms the prevalent PIT approach, 
and beats the FIFO strategy that relies on a strong bias on starting time. (2) We conducted a deep analysis of the serialization module and compared it with PIT and FIFO. 
This comparison effectively illustrates how the new approach functions.

\section{Methods}

\subsection{FIFO-SOT \& PIT-SOT}

SOT constructs the target labels for multi-talker speech by concatenating each speaker's labels according to specific serialization rules.
Suppose a multi-speaker speech signal $X$ involves speech from two different speakers, 
and these speakers appear sequentially in time, then the FIFO serialization strategy generates the target label sequence as follows:
 
\begin{equation}
\label{eq:seq}
L^{fifo} = [~L_1 \ \text{$<$sc$>$} \ L_2 \ \text{$<$sc$>$} \text{$<$eos$>$}~],
\end{equation}
\noindent where $L_1,\ L_2$ represent the content labels of the speech components 
of the two speakers, respectively. $<$sc$>$ represents `speaker change', 
and $<$eos$>$ represents `end of sentence'. Figure~\ref{fig:diag}(a) 
illustrates the FIFO strategy. The loss function for the FIFO-SOT with the target label $L^{fifo}$ is therefore defined as:

\[
Loss^{fifo} = \sum_{n} \text{CE}(y_n, L^{fifo}_n),
\]
\noindent where CE represents the cross-entropy loss and $L^{fifo}_n$ is the $n^{th}$ token in $L^{fifo}$, and $y_n$ is the $n^{th}$ output of the model.

For PIT, the order of $\{L_i\}$ in Eq.~(\ref{eq:seq}) can be permuted arbitrarily and the minimum CE loss resultant from these permutations is used as the loss for model training, formulated as:

\[
Loss^{pit} = \min_{\phi \in \Phi} \sum_{n} \text{CE}(y_n, L^{\phi}_n),
\]
\noindent where $\Phi$ represents the set of all possible permutations on $\{L_i\}$. Figure~\ref{fig:diag}(b) illustrates the computation of the PIT-SOT loss. 

\subsection{DOM-SOT}
\label{subsec:dom}

The proposed dominance-based SOT, abbreviated by DOM-SOT, 
is illustrated in Figure~\ref{fig:diag}(c). Overall, it is an AED architecture augmented by a 
serialization module (shown by the gray block); this module determines the order of the speakers when constructing the 
target label for model training, and the serialization module and the AED model are trained jointly. Note that once the model has been trained, the serialization module is discarded during inference. 

More specifically, given the multi-talker speech $X$, the feature sequence $h$ is derived by the AED 
encoder. $h$ is then forwarded to the serialization module, where the CTC losses are computed between 
$h$ and the labels of all speakers, i.e., $\{L_i\}$. According to these CTC 
losses, the speakers are arranged in ascending order, denoted by $\epsilon$, which is the output of the 
serialization module, and the target label $L^{\epsilon}$ is constructed according to $\epsilon$. Finally, the entire model is trained by:

\[
Loss^{dom} = \alpha * \min_{i=1,...,N}\{\text{CTC}(h,L_i)\} + (1-\alpha) * \text{CE}(y, L^{\epsilon}),
\]
\noindent where $N$ is the number of speakers in multi-talker speech. $\alpha$ is a hyper-parameter 
to tune the behavior of the encoder. Our experiments show that a small $\alpha$ is preferred, probably because it avoids the encoder concentrating on a single speaker and ignoring others.

\section{Experiment Settings}

\subsection{Training Data}
The official Train\_960h set from the LibriSpeech database~\cite{panayotov2015librispeech} was used 
to construct the training data. Each training sample could be clean speech, 2-talker speech, or 3-talker speech. These 
three types of speech samples are equal in quantity in the final training set. 
When constructing 2-talker or 3-talker samples, the clean utterances were randomly sampled from the Train\_960h set and then mixed using random and normalized weights.

To train the FIFO-SOT model, a 0.25s - 4s offset is imposed when mixing any two utterances. 
Without such offset, the model is hard to train as the bias on starting time is ambiguous. For the PIT-SOT and DOM-SOT models, 
the same offset occurs in 40\% of the training samples, while the remaining samples are mixed without any offset.

\subsection{Test Data}

We tested the various SOT models on both clean speech and multi-talker speech. 
The clean test used the \emph{test-clean} dataset from LibriSpeech. 
The 2-talker and 3-talker test samples were constructed by mixing the utterances 
listed in \emph{2mix-test-clean} and \emph{3mix-test-clean} from LibriMix~\cite{cosentino2020librimix}, 
respectively. These utterances are mixed with an offset of 0/1/2/3 seconds, to test performance in different scenarios. 

\subsection{Model Configuration}

All models in the experiments used 40-dimensional Mel-filter 
Fbank features as the input and adopted the same AED model structure where the 
encoder is a Conformer and the decoder is a Transformer. Before the encoder, two 
CNN layers perform sub-sampling. The kernel sizes and strides of both CNN layers were set to 3 
and 2. The first and second CNN layer's input/output channels were 1/256 and 256/256, respectively.

The encoder consisted of 9 Conformer blocks, each Conformer block consisted of a 4-head 
self-attention layer with 256 hidden units and two 2048-dimensional 
feed-forward layers (macaron style~\cite{lu2020understanding}). The decoder comprised 4 transformer blocks, 
each containing a 4-head 256-dimensional self-attention layer, a 256-dimensional cross-attention module, 
and a feed-forward linear layer with 2048 hidden units. For the DOM-SOT model, there is a serialization module. 
In our implementation, it was a simple linear layer that projects the encoder output to the word space, where the CTC loss was computed. The number of parameters of all the models is about 33M.

The output labels consisted of 5k wordpieces generated by 
the SentencePiece~\cite{kudo2018sentencepiece} toolkit\footnote[1]{https://github.com/google/sentencepiece.git} executed in 
the mode of the unigram language model. A blank label, an $<$\text{sc}$>$ token, and $<$\text{eos}$>$ token were also included in the label repository.

\subsection{Training Strategy}

The Adam optimizer was utilized for model training, employing a learning rate of 1e-3. 
All models were trained using a batch size of 32 over 60 epochs, with the initial 10 
epochs designated as the warm-up phase. Upon completion of training, the average of the checkpoints from the last 10 
epochs was selected as the final model. The hyper-parameter $\alpha$ in $Loss^{dom}$ was set to 0.1.

\subsection{Speaker-Aware WER}
\label{sec:method:sa}

The widely used naive WER is speaker-blind: it permutes the labels of all the speakers in the mixed speech and chooses the 
permutation that leads to the minimum WER by comparing the model output and the label sequence derived from the selected permutation. 
This naive WER tends to yield over-optimal results. For instance, if the placement of the $<$\text{sc}$>$ token in the output is 
inaccurately predicted or not predicted at all, the WER does not significantly change, but the performance could be substantially worse if we measure how the 
speech of \emph{each speaker} is recognized in average. 

This motivated the design of a speaker-aware WER. Specifically, the output of the model is firstly segmented into $n$ single-speaker 
segments (hypothesis) based on the $<$\text{sc}$>$ token, where $n$ represents the number of speakers predicted by the model. 
It is crucial to note that $n$ does not always correspond to the actual number of speakers, $m$, present in the mixed speech. 
A greedy approach is adopted to align the $n$ hypothesis segments and the $m$ references. For each reference, the best-matched 
hypothesis is discovered in terms of WER, and the matched reference and hypothesis are removed from the reference set and the 
hypothesis set respectively. The match continues until either the reference set or the hypothesis set is empty. Ultimately, 
each unpaired reference is assigned an empty hypothesis, and each unpaired segment is assigned an empty reference. After this alignment, insertions, 
deletions, and substitutions are calculated with each reference/hypothesis pair, and the speaker-aware WER is computed simply by 
computing the ratio of these errors among all the tokens in the target label.

\[
\text{WER} = \frac{\sum_{j=1}^{max(n,m)}(S_{j} + I_{j} + D_{j})}{\sum_{j=1}^{max(n,m)}{C_{j}}},
\]

\noindent where $j$ indexes the hypothesis/reference pair, $S_j$, $I_j$, $D_j$ are the number of substitutions, 
insertions, and deletions computed from the $j$-th pair, and $C_j$ is the number of tokens of the $j$-th reference segment. 

%\vspace{-1mm}
\section{Experimental Result}

We compare the three SOT systems constructed with different serialization strategies: FIFO-SOT, PIT-SOT, and DOM-SOT. 
The test includes three groups: clean test, 2-mix talker test, and 3-mix talker test. In each of 
the mix-talker tests, 4 test conditions are designed, according to the offset among speakers. 
Note that the models are the same for all these 3 groups and 4 conditions. We report both speaker-blind WERs and speaker-aware WERs.

\begin{table*}[htpb!]
    \caption{Performance of SOT models with different serialization strategies.}
    \vspace{-2.5mm}
    \label{tab:wer}
    \centering
    \scalebox{0.96}{
    \begin{tabular}{l|c|cccc|cccc}
     \toprule
    &\multicolumn{1}{c}{Clean}& \multicolumn{4}{|c}{2-Mix}  & \multicolumn{4}{|c}{3-Mix} \\
    \cmidrule(r){2-2}\cmidrule(r){3-6} \cmidrule(r){7-10} %\cmidrule(r){2-5} \cmidrule(r){6-9}
    Offset & - &   3s        &  2s & 1s & 0s &  3s & 2s &  1s & 0s \\
    \midrule
     \multicolumn{10}{c}{Speaker-Blind WER (\%)}\\
     \midrule
     FIFO-SOT                 & 5.36           & 5.94        & 6.41          & 7.00         & 14.87 & 9.96              & 11.65          & 16.53          & 42.66 \\
     PIT-SOT                  & 5.30           & 6.52          & 6.74          & 7.09         & 8.95           & 11.82           & 12.76          & 16.89          & 22.19 \\
     DOM-SOT          & \textbf{5.17}     & \textbf{5.56}   & \textbf{5.82} & \textbf{6.06} & \textbf{7.11}  & \textbf{9.96}     & \textbf{11.44}  & \textbf{15.02} & \textbf{19.49} \\
    \midrule
    \multicolumn{10}{c}{Speaker-Aware WER (\%)}\\
    \midrule
    FIFO-SOT                 & 5.69   & \textbf{6.26}         & 6.58          & 7.22         & 21.53 & \textbf{14.96}  & \textbf{13.06}          & 17.58          & 53.68 \\
    PIT-SOT                  & 5.48           & 8.05          & 6.95          & 7.29         & 9.23           & 21.17           & 17.38          & 17.99          & 23.26 \\
    DOM-SOT          & \textbf{5.25}          & 6.93   & \textbf{6.08} & \textbf{6.28} & \textbf{7.31}        & 15.74           & 14.78  & \textbf{16.01} & \textbf{20.51} \\
     \bottomrule
     \end{tabular}}
    % \vspace{-0.15mm}
 \end{table*}

\subsection{FIFO-SOT vs PIT-SOT}

The comparison between FIFO-SOT and PIT-SOT aligns closely with our expectations. The performance difference between the two models on clean 
data is insignificant. Moreover, in the 2-mix and 3-mix groups, FIFO-SOT consistently outperforms PIT-SOT when the offset is not 0s. This is not surprising as 
the bias on starting time is clear in these test conditions, and the 
bias has been explicitly exposed to the model when training the FIFO-SOT model. This 
also aligns with the results reported in~\cite{lu2021streaming, kanda2020serialized}. However, when 
the offset is 0s, the performance of FIFO-SOT drastically declines. It is noteworthy that the true offset of the 
speakers in the 0s-offset condition is not exactly 0 due to the leading silence within each utterance. However, this slight 
offset is not sufficient to form a clear bias, resulting in poor performance. This is an illuminating result and indicates that if 
the bias is ambiguous in the test, SOT models relying on that bias may completely fail. This is perhaps a general rule and may apply to any bias.

In comparison, PIT-SOT remains relatively stable in all the test conditions: it is slightly 
worse than FIFO-SOT when the bias on time is clear, but remains a rather good performance when the bias disappears. This underscores its effectiveness in dealing with multi-talker ASR tasks. 

\subsection{DOM-SOT vs PIT-SOT and FIFO-SOT}

First of all, DOM-SOT outperforms PIT-SOT under all testing conditions. This is a 
highly promising result, as the two models were trained with identical resources and nearly identical 
structures (the extra parameters from the additional projection layer of the serialization module in DOM-SOT can be largely ignored). This indicates 
that the performance gain obtained with DOM-SOT is of no cost. We therefore conclude that DOM-SOT, like PIT-SOT, is a general architecture for dealing with multi-talker speech, and it is more powerful than PIT-SOT.

When comparing with FIFO-SOT, we can find that in the 0s-offset group, DOM-SOT shows an overwhelming advantage. 
Even on the conditions with a longer offset for which the bias is very clear and the training and test conditions match perfectly for 
FIFO-SOT, DOM-SOT can achieve comparable or even better performance compared to FIFO-SOT. We highlight that in these conditions, PIT-SOT never beats FIFO-SOT. 
This provides strong evidence that DOM-SOT is a more powerful architecture than PIT-SOT in tackling multi-talker speech. 

\subsection{Speaker Aware vs Speaker Blind}

Table~\ref{tab:wer} presents both speaker-blind WER and speaker-aware WER. 
It is evident that the speaker-aware WER is a more conservative metric than the speaker-blind WER, as discussed in Section~\ref{sec:method:sa}.
Moreover, the comparison between the two types of WERs reveals extra information about the model behavior. 
For instance, DOM-SOT outperforms in all the conditions in terms of speaker-blind WER but is slightly worse than FIFO-SOT in the 
3s-offset conditions if the metric is speaker-aware WER. This implies that DOM-SOT produces better target labels overall but the location of the $<$sc$>$ token is not predicted as well as FIFO-SOT. 

%\vspace{-1.5mm}
\section{Analysis \& Discussion}
\label{ana}

In this section, we aim to discover how DOM-SOT works and outperforms PIT-SOT and FIFO-SOT. 
To achieve this, we conducted an analysis of the dominance scores of speech components within mixed speech samples, 
i.e., the CTC loss computed by the serialization module with the transcript of each speech component. 
The mixed samples are from the \textbf{\textit{2-mix}} group, under the \textbf{\textit{0s-offset}} condition and the \textbf{\textit{3s-offset}} condition. 

First of all, we found that nearly all samples adhered to the serialization rule based on dominance scores. 
More specifically, for the 2-mix data with 0s/3s offsets, 99.7\% and 98.6\% of the test data, respectively, were transcribed by putting the speech component with a lower CTC loss ahead. 
This observation indicates that \textbf{the DOM-SOT model has effectively learned to utilize CTC loss as a criterion for organizing its output sequence.}

\begin{figure}[ht]
     \centering
      \includegraphics[width=1\linewidth]{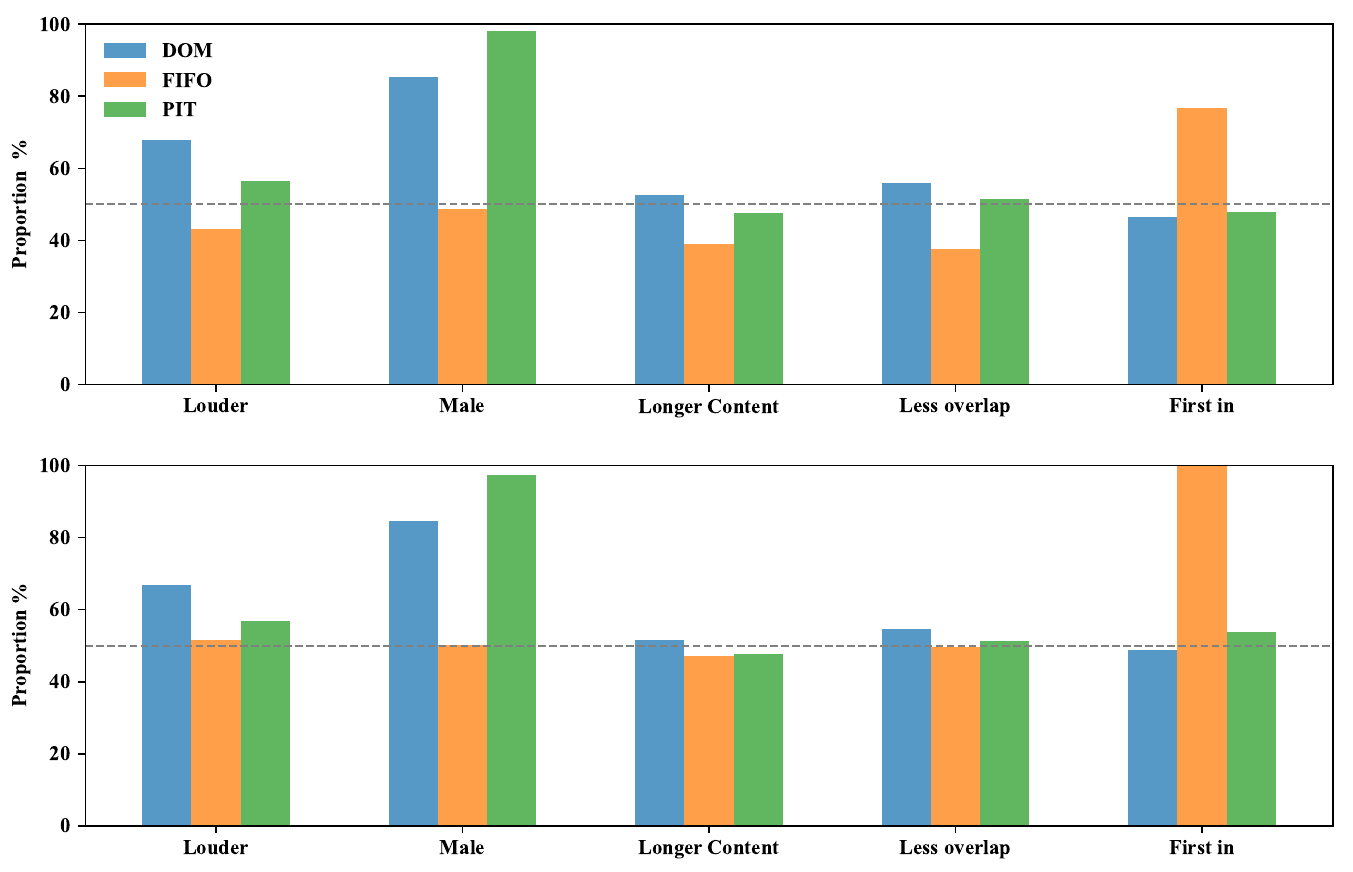}
     \caption{Proportion of factors contributing to the dominance of dominant speech under the condition 2-mix with 0s/3s offset.}
     \label{fig:ana}
     \vspace{-1.0mm}
\end{figure}

%which stateLoudness, Content length, and Clear cue length are closely linked to the dominance factor utilized by humans. 
%We used the MFA tool\footnote{https://mfa-models.readthedocs.io} to annotate precise pronunciation locations in clean-test data, thereby acquiring the most accurate statistical values.

Next, we attempt to identify commonalities in dominant speech. We investigated five potential 
factors: Loudness, Gender, Content length, Length being overlapped, and Start time. These factors were 
selected based on human perception of dominant speech~\cite{kawata2020neural, mast2002dominance, jie2010recognize}. 
By comparing dominant and non-dominant speech across these factors — for instance, if the dominant speech is louder than 
the non-dominant speech — one can identify potential biases used to determine dominant speech. To ensure the results are meaningful, 
the test sets were filtered to maintain balance across all factors. We also conducted a similar analysis for FIFO-SOT and PIT-SOT. 
Since the concept of `dominant speech' is irrelevant in these methods, we define the first transcribed speech as dominant. 
This is acceptable as the comparison between the two speech components is what matters.

Figure~\ref{fig:ana} displays the results. It is evident that for DOM-SOT, there is a distinct bias on loudness and gender. 
This suggests that when two speeches are mixed, the encoder tends to prioritize recognizing louder male speech as dominant. 
In contrast, FIFO-SOT uses starting time as the bias, while PIT-SOT uses gender as the major bias. 
Such findings indicate that DOM-SOT employs a multi-bias approach, making it more robust than single-bias approaches. 
This clarifies why it outperforms FIFO-SOT and PIT-SOT. Importantly, the biases in DOM-SOT are identified by the model itself rather than 
being predefined by human intuition. Note that the biases shown with DOM-SOT and PIT-SOT are stable in both the 0s-offset and 3s-offset conditions, 
indicating that they are truly used by the model rather than from statistical randomness. 

\section{Conclusion \& Future Work}

We propose a model-based serialization strategy for the SOT framework, named DOM-SOT. 
This strategy augments the standard AED model with a serialization module that computes CTC loss for each speech component and orders the labels according to this loss.
Surprisingly, this approach demonstrated a remarkable advantage over the prevalent PIT approach in our multi-talker ASR experiments, 
and yielded comparable or superior results to methods based on human-defined biases, like FIFO, even in test conditions where the bias is strong.
Deep analysis showed that the serialization module has discovered multiple factors to order the speech components, especially loudness, and gender. This multi-factor nature explains its superior performance. 

We also advocate the use of speaker-aware WER as the metric for measuring multi-talker ASR performance, as 
it highlights the importance of the speaker-change token. Analysis has shown that the performance of all SOT systems decreases when 
measured by speaker-aware WER, indicating potential errors in the speaker-change token, in particular with the proposed strategy. Future 
work will investigate this problem, and conduct extensive experiments on other datasets to confirm the superiority of the dominance-based approach over the PIT approach.

%\section{Acknowledgements}
% \ifinterspeechfinal
% %     The Interspeech 2024 organisers
% \else
% %     The authors
% \fi

\bibliographystyle{IEEEtran}
\bibliography{mybib}

\end{document}